\documentclass[conference]{IEEEtran}
\IEEEoverridecommandlockouts
\usepackage{cite}
\usepackage{amsmath,amssymb,amsfonts}
\usepackage{algorithmic}
\usepackage{graphicx}
\usepackage{textcomp}
\usepackage{xcolor}
\usepackage{subfigure}

\usepackage{epstopdf}
\usepackage{color}
\usepackage{bm}
\usepackage{amsmath}
\usepackage{mathtools}
\usepackage{adjustbox}
\usepackage[font={small}]{caption}
\usepackage{mathrsfs}
\usepackage{placeins}
\usepackage{float}

\title{Methods and Results for Quantum  Pulse Control on Superconducting  Systems\\
\thanks{
This work was supported by the Laboratory Directed Research and Development Program (\#19-002) of Brookhaven National Laboratory, which is operated and managed for the U.S. Department of Energy Office of Science by Brookhaven Science Associates under contract No.DE-SC0012704.\\
{\footnotesize \textsuperscript{*}Current affiliation: NVIDIA Corporation}.}
}

\author{\IEEEauthorblockN{ Elisha Siddiqui Matekole}
\IEEEauthorblockA{\textit{Computational Science Initiative} \\
\textit{Brookhaven National Laboratory}\\
Upton, New York, USA \\
esiddiqui@bnl.gov}
\and
\IEEEauthorblockN{Yao-Lung Leo Fang*}
\IEEEauthorblockA{\textit{Computational Science Initiative} \\
\textit{Brookhaven National Laboratory}\\
Upton, New York, USA}
\and
\IEEEauthorblockN{ Meifeng Lin}
\IEEEauthorblockA{\textit{Computational Science Initiative} \\
\textit{Brookhaven National Laboratory}\\
Upton, New York, USA} 
}
\begin{document}
\maketitle
\begin{abstract}
The effective use of current Noisy Intermediate-Scale Quantum (NISQ) devices is often limited by the noise which is caused by interaction with the environment and affects the fidelity of quantum gates. In transmon qubit systems, the quantum gate fidelity can be improved by applying control pulses that can minimize the effects of the environmental noise. In this work,  we employ physics-guided quantum optimal control strategies to design optimal pulses driving quantum gates on superconducting qubit systems. We test our results by conducting experiments on the IBM quantum hardware using their OpenPulse API. We compare the performance of our pulse-optimized quantum gates against the default quantum gates and show that the optimized pulses improve the fidelity of the quantum gates, in particular the single-qubit gates. We discuss the challenges we encountered in our work and point to possible future improvements.   
\end{abstract}

\section{Introduction}
In the current era of noisy intermediate-scale quantum (NISQ) regime, we expect to see quantum computers with hundreds or thousands of imperfect qubits. Quantum technology companies have made strides towards quantum computers with increasing numbers of qubits. Recently, IBM broke the 100 qubit barrier with a 127 qubit quantum processor, which is a big step towards practical quantum computation. In addition to increasing the number of qubits it is also essential to develop high-fidelity quantum gates to demonstrate the quantum advantage and achieve fault-tolerant quantum computation. However, the current quantum hardware is sensitive to noise and the quantum logic gates often suffer from errors that reduce the fidelity of the respective quantum operations. This in turn affects the ability to reliably carry out large-scale quantum computations. Therefore, there is a need to develop techniques that can improve the quantum gate fidelity as close as possible to unity, and improve the performance of quantum hardware. 
Previously, Lie group theory and geometric formalism has been implemented for dynamically correcting gates in single-, and multi-qubit systems \cite{Barnes2020}. Other strategies used for controlling quantum systems are Lyapunov and bang-bang methods \cite{ESW,CONG20209220,Bhole2015}. However, these techniques have a high numerical cost. 
In recent years, optimal-control-based methods have been widely applied to quantum systems, to maximize the fidelity of the quantum processes that drive the qubit from one state to another. Generally, in  optimal control, for a given model of the quantum system such as transmons, ions and neutral atoms etc., the control pulses that minimize the associated cost function are obtained.
Some of the earliest optimal-control approaches are Krotov \cite{Krotov} and gradient ascent pulse engineering (GRAPE) \cite{GRAPE1}. Of late, methods such as chopped random basis optimization (CRAB) \cite{CRAB} and gradient optimization of analytic controls (GOAT) \cite{GOAT} have been proposed. There are several open-source software packages such as \texttt{C3} \cite{c3-optimize} and \texttt{qopt} \cite{qopt2021} that also utilize optimal control of pulses. So far these methods have been implemented for numerical simulations and not been demonstrated on real quantum hardware. Recently, researchers have been looking into leveraging deep reinforcement learning (DRL) techniques to prepare these target quantum gates from any initial state, and are robust to errors \cite{He2021, qctrl2021}. The DRL algorithm in \cite{qctrl2021} was executed on the experimental quantum hardware of IBM.  

One of the most promising technology to create a full scale quantum computer is through superconducting devices \cite{supqubit2017}. There are several realizations of superconducting qubits~\cite{Krantz2019}. We explore the impact of quantum optimal control on the gate performance of IBM quantum systems by executing the custom piece-wise-constant (PWC) pulses directly on a superconducting quantum computer using OpenPulse API \cite{Qiskitpulse}.

This manuscript is organized as follows. First we provide a brief overview of quantum optimal control and its implementation in QuTiP \cite{JOHANSSON_QuTiP}. Second, we outline the implementation of the offline pulses on the real IBM Q hardware. We compare the error-rates of our optimized gates with the default device-gates. We end with a discussion on the challenges we faced and future outlook.

\section{Background}
Quantum optimal control (QOC) is an important tool used extensively in Physics and Chemistry applications where the goal is to steer the time evolution of the quantum system to a particular target state, unitary operation or a desirable state-to-state transfer. It synthesizes control fields for a particular control target, constraints and time evolution of quantum system. The time-dependent Hamiltonian can be written as $H(t)=H_{0}+\sum_{i=1}^{n} u_{i}(t)H_{i}$, where the first term is the drift Hamiltonian and second term represents the control Hamiltonian, and $u_{i}(t)$ are the control functions that describe the strength with which, each of the control Hamiltonian acts as a function of time. The goal is to determine a set of $\{u_{i}(t)\}$, for each control, to optimize the relevant cost function $\mathscr{C}[\{u_{i}(t)\}]$. 
We use QOC method to design the optimal pulses to execute the quantum gate on the quantum computer. The cost function for this particular case is gate infidelity $\mathscr{C}=1-\mathscr{F}=1-\frac{1}{\mathscr{N}}|\mathrm{Tr}(U_{\mathrm{t}}^{\dagger}U_{\mathrm{f}})|^{2}$, where $U_{\mathrm{t}}$ is the target unitary gate, and $U_{\mathrm{f}}$ is the unitary realized by the control pulses. 

Typically, the control problem cannot be solved analytically. Hence one resorts to numerical methods. The main quantum optimal methods used to minimize the cost function are gradient based approaches. One of the oldest techniques is GRAPE (Gradient Ascent Pulse Engineering)\cite{GRAPE1}. However, this method converges very slowly to the optimal cost function. Also, control pulses based on the CRAB method can be derived from truncated Fourier series. However,  the CRAB algorithm utilizes a direct search approach which makes the convergence very slow even for small optimization variables \cite{CRAB,Riaz2019}.
A second-order GRAPE method known as L-BFGS-B is a limited-memory algorithm which, as the name suggests, requires much less memory than its precursor and converges faster \cite{GRAPE2}. We have also tested another optimization method called Simultaneous Perturbation Stochastic Approximation (SPSA)\cite{SPSA}. It is a gradient approximation, but unlike GRAPE it does not measure the gradient of the cost function but the cost function itself. The gradient approximation measures the objective function at only two points. We found that L-BFGS-B converges faster and gives much smaller fidelity error than SPSA. Therefore we will utilize L-BFGS-B as our numerical optimization method of choice.

\section{Implementation}
\label{sec:impl}

\subsection{Pulse Optimization with QuTiP}
\begin{figure}[t]
    \centering
    \includegraphics[width=0.7\columnwidth]{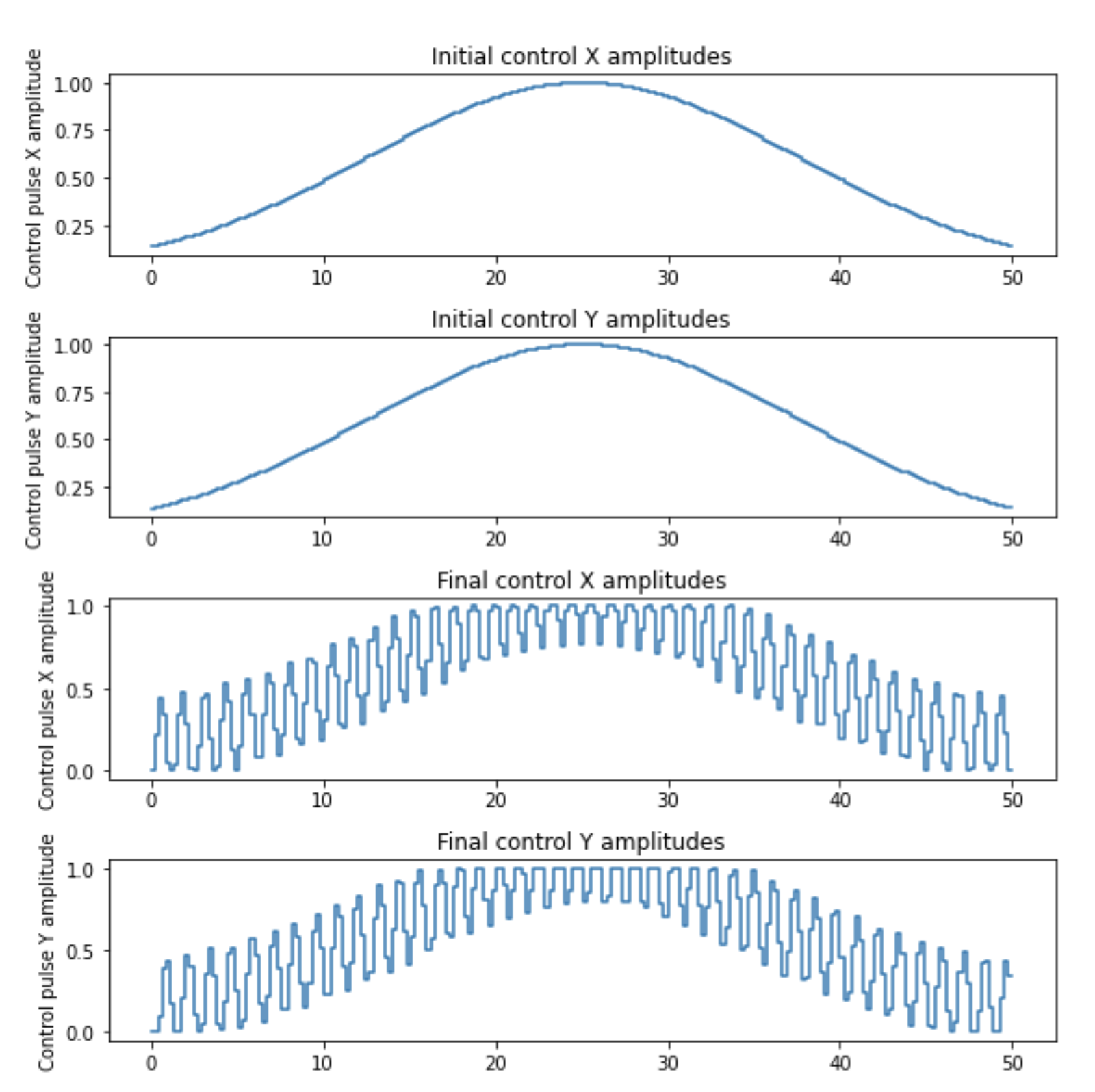}
    \caption{An example of control pulses for X gate generated by the pulseoptim function in QuTiP. The top two panels represent the initial pulses for each control term and the lower two panels show the output pulses after the optimization.}
    \label{fig:x_qutip}
\end{figure}

In this work, we generate optimal pulses using the QuTiP library~\cite{JOHANSSON_QuTiP}. QuTiP is an open-source Python package to simulate dynamics of quantum systems, and provides useful tools we need for the pulse optimization. In particular, we use the QuTiP {\tt pulseoptim} function, which employs the L-BFGS-B as the optimization algorithm \cite{JOHANSSON_QuTiP} and outputs the optimized pulse coefficients for each control term. Our current target quantum architecture is the IBM Q superconducting qubits, since we only have access to them at the time of this work. The Hamiltonian that represents the IBM Q hardware can be constructed based on the information provided by IBM, such as the coupling between qubits and qubit frequencies. The Hamiltonian is then used to generate the control pulses through QuTiP {\tt pulseoptim}.  An example control pulse generated this way for the X gate is given in Fig.~\ref{fig:x_qutip}. 

\subsection{Pulse Implementation with Qiskit}
We then test the optimized control pulses on the real hardware to see if the quantum gate fidelity has been improved. To do this, we employ the IBM Qiskit-Pulse library, which is a low-level quantum programming tool using pulses\cite{Qiskitpulse}. The Qiskit-Pulse is a front-end implementation of the OpenPulse interface~\cite{McKay2018} which allows the user to modify the pulse parameters and translate the pulse program to an executable circuit which can be implemented on the quantum hardware. 
The Qiskit-Pulse consists of a pulse shape library, pulse channels, schedules and instructions. The user can input pulse shape, duration, amplitude using the pulse waveform function or by calling any of the existing shapes such as $drag$ (Derivative Removal by Adiabatic Gate) pulse in the pulse library. These variables can be modified to customize the pulse driving any quantum gate.

\begin{figure}[htbp]
    \centering
    \includegraphics[width=0.9\linewidth]{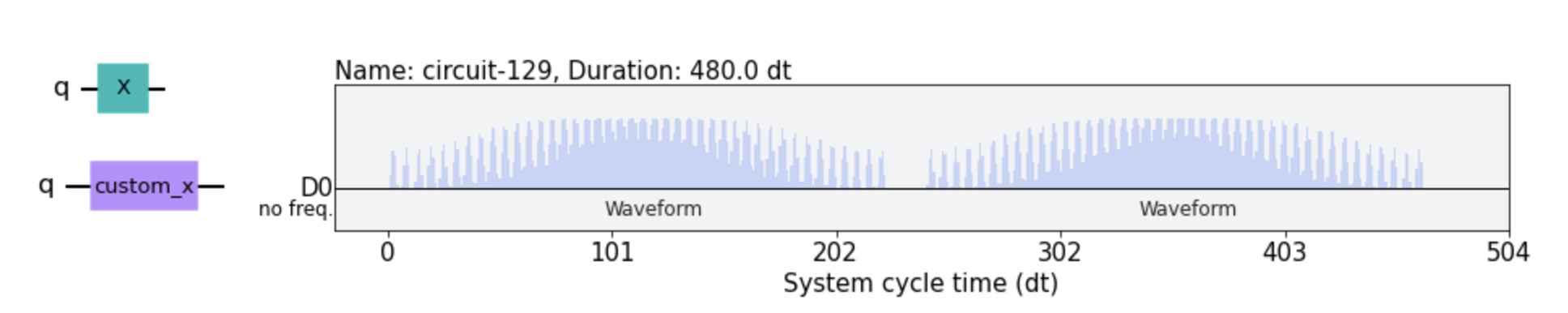}
    \caption{(Color online) The control pulse implemented on $ibmq\_montreal$ device. Here we use $drag$ pulse shape. 
    D0 is the DriveChannel for qubit 0. 
    The default X gate is replaced by our optimized X gate, which is confirmed in the transpiling process.}
    \label{fig:X_pulse}
\end{figure}

The pulse is built/implemented using the {\tt pulse.build} function, where the pulse is assigned to the respective channels. 
The channels available are \texttt{DriveChannel}, \texttt{ControlChannel} and \texttt{AcquireChannel}. In this work we use the first two channels. The \texttt{DriveChannel}, as the name suggests, drives the associated qubit with the input pulse. The \texttt{ControlChannel} is similar to the \texttt{DriveChannel}, but it is implemented in the case of multi-qubit gates. 

The pulse instruction can be converted into a gate-level instruction using the {\tt Gate} function in Qiskit, but before that it needs to be mapped to the backend using the instruction schedule map. The user can also view whether the pulse has been successfully mapped. The pulse gate can now be implemented at the circuit level. One can view the underlying pulses of the quantum circuit using the {\tt schedule} function in Qiskit. The circuit is executed on the hardware using this schedule. An example of the pulse schedule for the X gate as implemented on the $ibm\_montreal$ device is shown in Fig.~\ref{fig:X_pulse}.

\subsection{Benchmarking}
To test how well our optimized pulses work, we use two different measures. First, we check the probability distribution of the qubit output state after the gate operation, and see if it is consistent with the expected probability distribution for state $|0\rangle$ or $|1\rangle$. Next, we check the average gate errors using randomized benchmarking (RB)~\cite{RB}, which is implemented in Qiskit. However, the standard RB procedure in Qiskit does not allow the inclusion of custom gates. Instead, we use the interleaved randomized benchmarking (IRB), which allows us to use the custom gates with the optimized pulses. For an in-depth discussion on IRB please see \cite{IRB} and Qiskit documentation page. We will explain in more detail how these benchmarks are used in our experiments in Section~\ref{sec:results}. 

\section{Experiments On IBM Q }
\label{sec:results}

In the following sections, we present the results for implementing our control pulses using the procedure described in Section~\ref{sec:impl}. After a discussion of the details of the IBM Q test systems, we will present our results for the X (not) gate, the $\sqrt{x}$ (square root not) gate, H (Hadamard) gate and the two-qubit CNOT (control not or CX) gate. 

\subsection{Details of the IBM Q Test Systems}
Our experiments were conducted on $ibmq\_toronto$ and $ibmq\_montreal$ systems. The $ibmq\_toronto$ system has a quantum volume of $32$ with $27$ qubits. The average $T_{1}$ is $83.52~\mu$s. We utilize qubit $0$ for our experiments, which has a frequency of $5.225$ GHz and average single qubit gate error of $3.068\times10^{-4}$. For $ibmq\_montreal$ the quantum volume is $128$ with $27$ qubits. The average $T_{1}$ is $86.76~\mu$s. Here we again utilize qubit $0$ for our experiments, which has a frequency of $4.911$ GHz and average single qubit gate error of $4.268\times10^{-4}$. We chose qubit $0$ for both devices as it is connected to only qubit $1$. This simplifies the implementation of the Hamiltonian model numerically. We would also like to mention that both these devices have the same topology \cite{IBMQ}.

\subsection{X Gate}

\begin{figure}[t]
\centering

   \includegraphics[width=0.75\columnwidth]{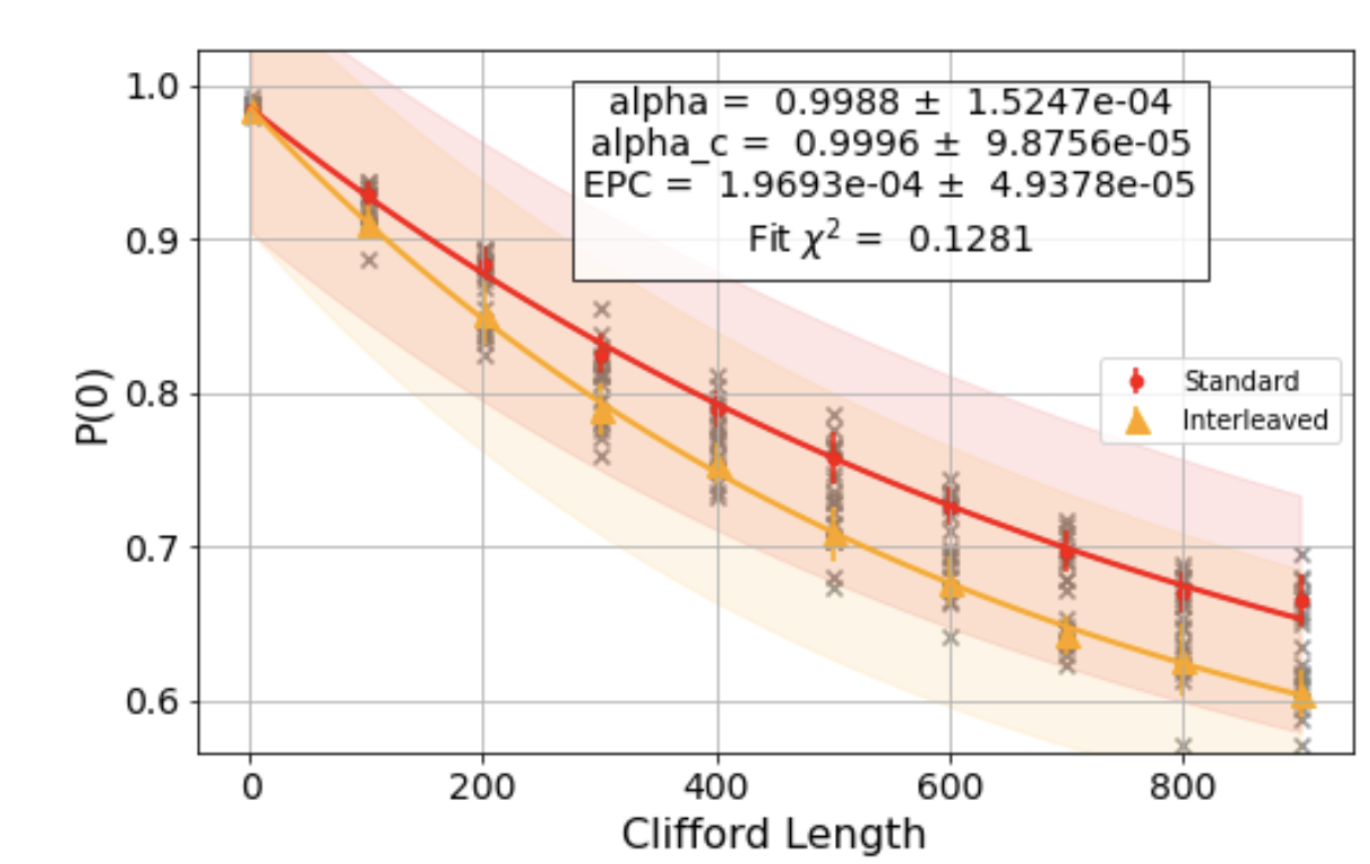}
   \includegraphics[width=0.75\columnwidth]{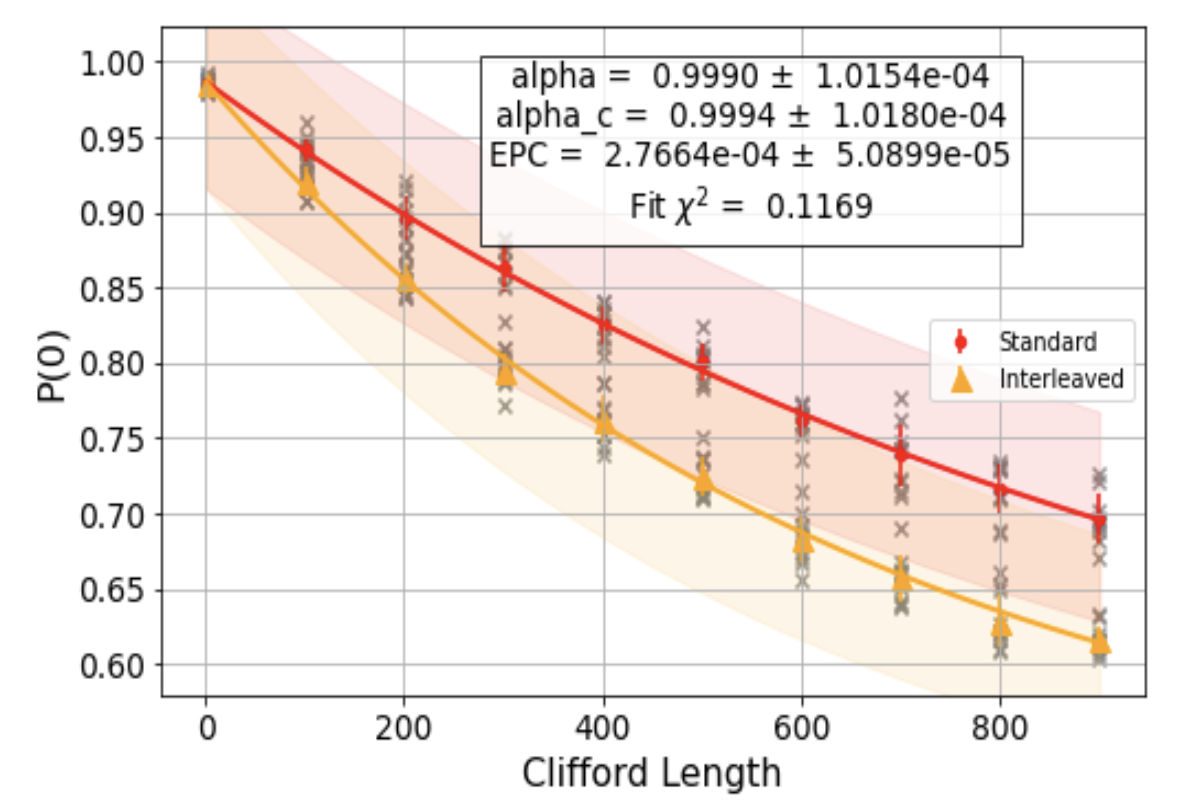}
   \includegraphics[width=0.75\columnwidth]{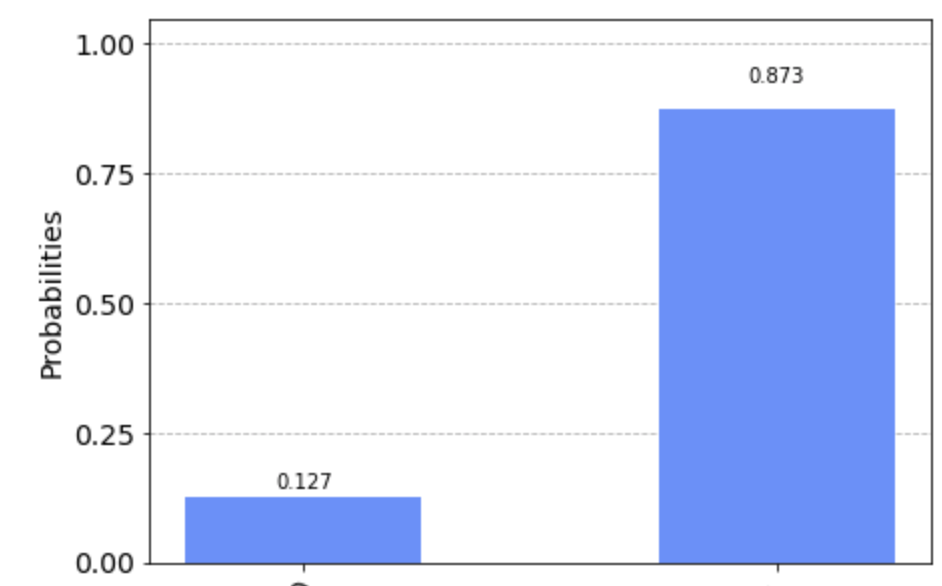}
\caption{(Color online) (Top) IRB result for the custom X gate obtained from optimized pulse controls. The error rate here is $(2.0\pm0.5)\times10^{-4}$. (Middle) IRB result for the original (default) X gate as implemented in Qiskit. The error rate here is $(2.8\pm0.5)\times10^{-4}$. (Bottom) The probability distribution of the qubit state measurements.}
\label{fig:IRBX}
\end{figure}

First we discuss the results for the X gate, which constitutes one of the basis gates in Qiskit. It is also known as a $\pi$-pulse gate. In IBM Q systems the $\pi$-pulse is regularly calibrated by carrying out a Rabi experiment. More information on this can be found here \cite{Qiskit-Textbook}.

We use the duffing oscillator Hamiltonian for the system, and Pauli X, Pauli Y are the control terms for the control Hamiltonian. We import the values of qubit frequency and decoherence rate from the backend description provided by IBM. The initial pulse shape is chosen to be $drag$, the amplitude bound is $[0,1]$. The evolution time for each control term is $52$ ns and the total pulse duration is $480$ dt $\sim$ $105$ ns. The optimized pulse coefficients are obtained using the QuTiP optimal pulse module. The pulses obtained in QuTiP and the control pulse implemented on $ibmq\_montreal$ are shown in Fig.~\ref{fig:X_pulse}.
\begin{figure}[t]
\centering
   \includegraphics[width=0.75\columnwidth]{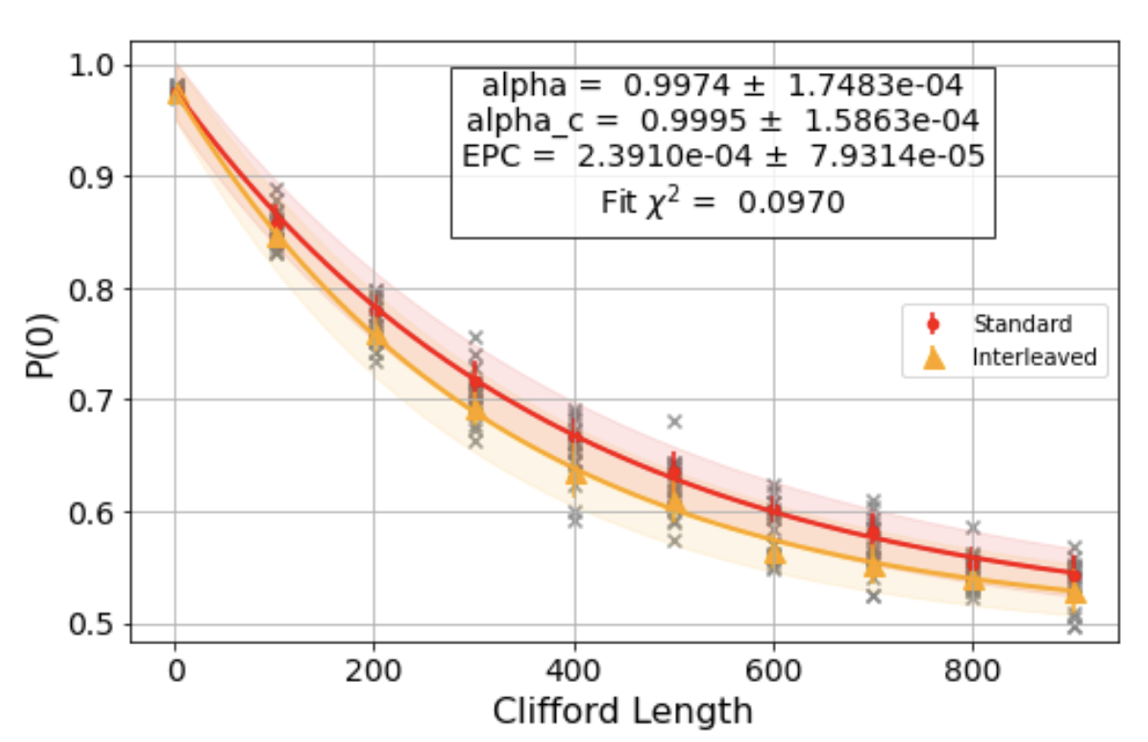}
   \includegraphics[width=0.75\columnwidth]{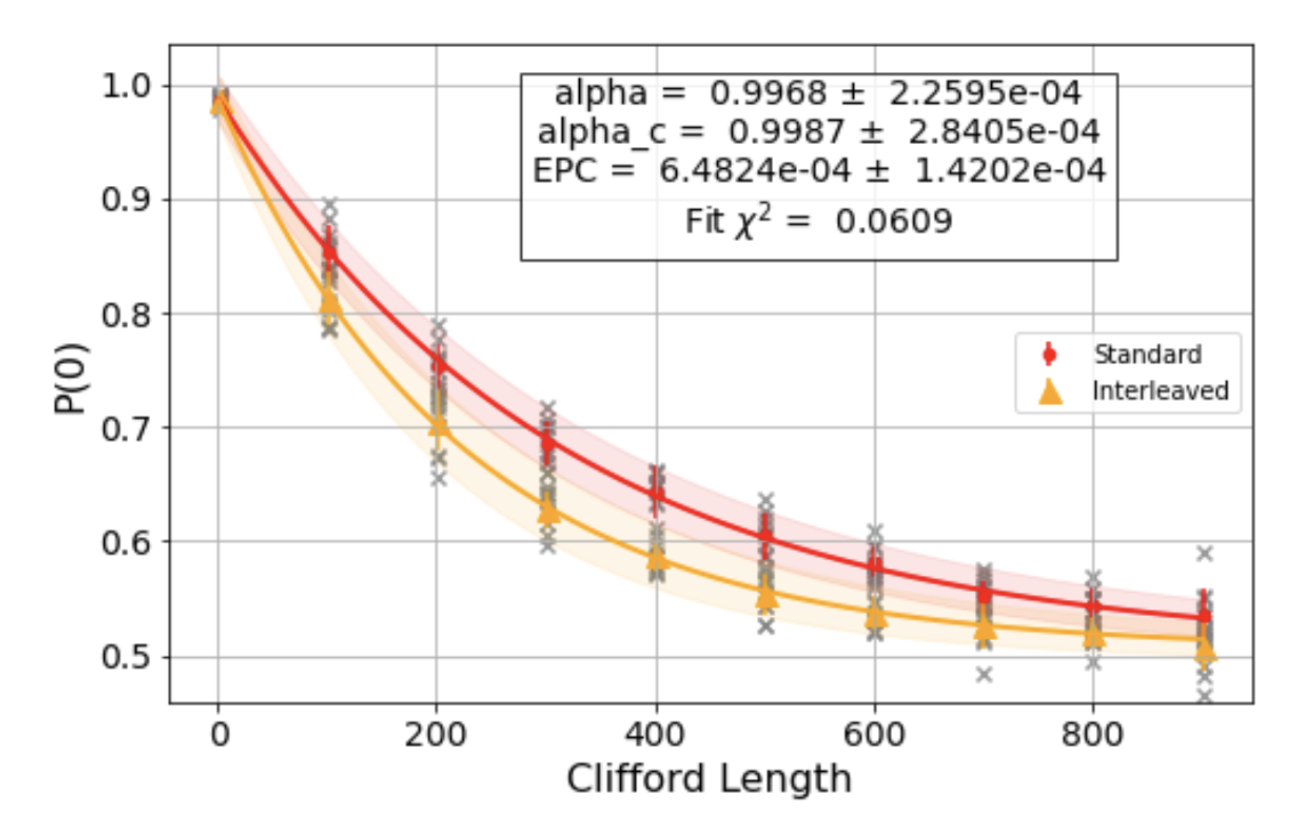}
   \includegraphics[width=0.75\columnwidth]{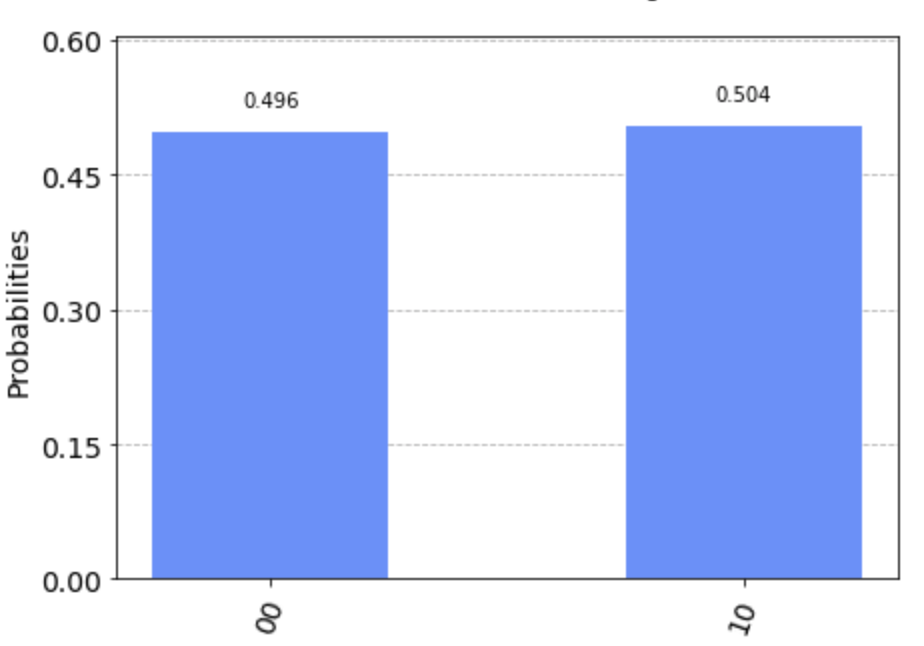}
\caption{(Color online) (Top) IRB result for the custom $\sqrt{x}$ gate obtained from optimized pulse controls. The error rate here is $(2.4\pm0.8)\times10^{-4}$. (Middle) IRB result for the original (default) $\sqrt{x}$ gate as implemented in Qiskit. The error rate here is  $(6.5\pm1.4)\times 10^{-4}$.  (Bottom) Probability distribution of the qubit state after applying the pulse-optimized $\sqrt{x}$ gate. }
\label{fig:IRBSX}
\end{figure}
We test our pulses on the $ibmq\_montreal$ backend. We first create a circuit to prepare and measure a NOT gate, then reduce it to a pulse schedule before running the pulse job on the device. We measure the qubits, and plot the probability distribution for $|0\rangle$ and $|1\rangle$ state measurements in the bottom panel of Fig.~\ref{fig:IRBX}. We can see that the output state has $87.3\%$ probability of being in state $|1\rangle$ (up to measurement errors) after implementing the custom X gate in the quantum circuit, while an error-free gate would result in 100\% probability.

Next, we characterize our gate using IRB, which gives the estimate of the average error-rate of our custom gate. The IRB experiment generates both the standard RB sequence and the interleaved one. This experiment calculates the probabilities to get back to the ground state, fits the probability curve to get $\alpha$ (depolarizing parameter) and $\alpha_{c}$ (ratio of depolarizing parameter of IRB to RB curve) and calculates the interleaved gate error. 
In the top and middle panels of Fig.~\ref{fig:IRBX} we show the error rates of the custom X gate and the default X gate, using IRB in Qiskit. We find that the error rate for our pulse optimized gate is about $28\%$ lower than the default X gate implemented in Qiskit. 
\subsection{Square root NOT ($\mathbf{\sqrt{x}}$) gate}
\begin{figure}[t]
\centering
   \includegraphics[width=0.75\columnwidth]{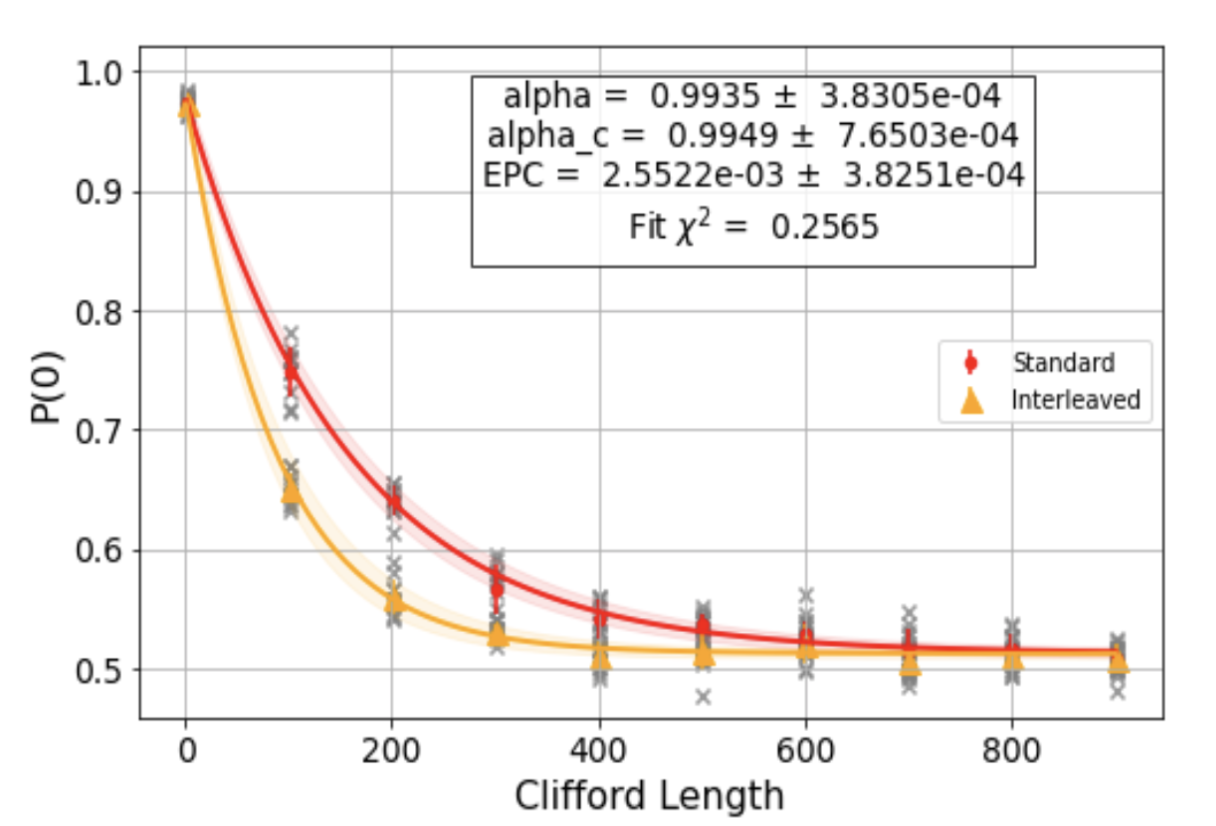}
   \includegraphics[width=0.75\columnwidth]{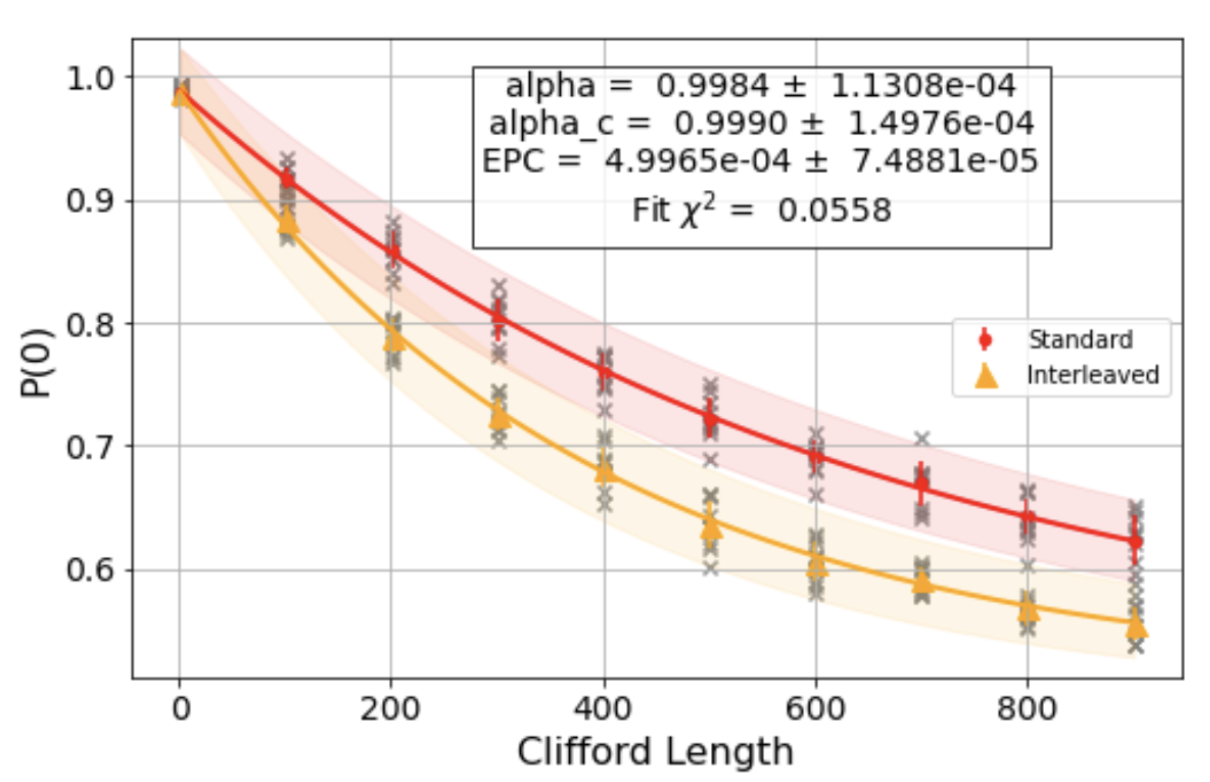}
   \includegraphics[width=0.75\columnwidth]{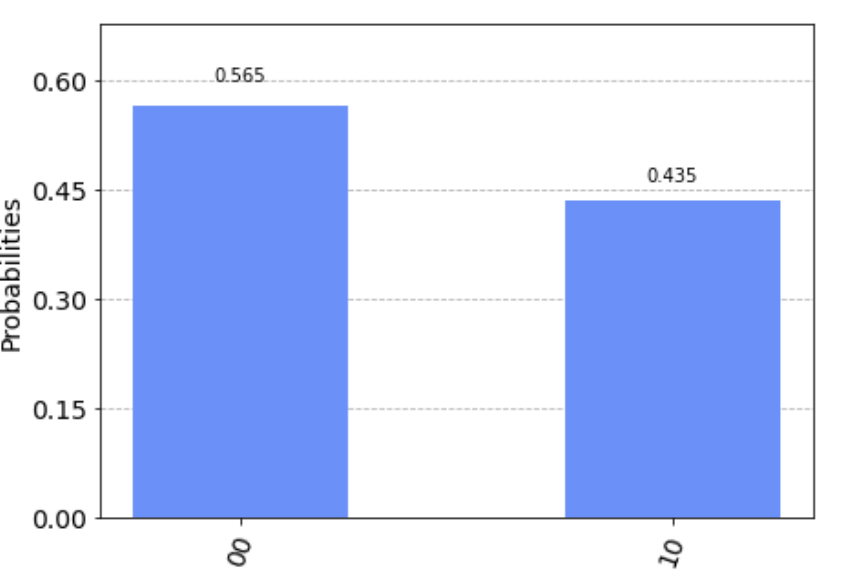}
\caption{(Color online) (Top) IRB result for the custom H gate obtained from optimized pulse controls. The error rate here is  $(2.6\pm0.4)\times10^{-3}$. (Middle) IRB result for the original (default) H gate as implemented in Qiskit. The error rate here is  $(5.0\pm0.7)\times10^{-4}$. (Bottom) Probability distribution of the qubit state after applying the pulse-optimized Hadamard gate operation.  }
\label{fig:IRBH}
\end{figure}
Square-root-not gate ($\sqrt{x}$) is also a basis gate in IBM Q systems.
Similar to the X gate, we use the duffing oscillator Hamiltonian with Pauli X as the control term for the control Hamiltonian. The optimized pulses are obtained using QuTiP optimal pulse module. The control pulse of length $736$ dt $\sim$ $162$ ns is implemented on $ibmq\_montreal$.

For open systems the presence of dissipative terms can prevent or enable certain states to be reached. We found that for the $\sqrt{x}$ operation we were not able to reach a global minimum of the cost function. Hence, for the case of $\sqrt{x}$ gate we neglected the decoherence processes during the optimization for computational simplicity.   

The error rate for the custom $\sqrt{x}$ gate and the default $\sqrt{x}$ gate is obtained from the results of IRB as shown in Fig.~\ref{fig:IRBSX}. We observe that the error rate of our custom $\sqrt{x}$ gate is $63\%$ lower than the default gate. The bottom panel of Fig.~\ref{fig:IRBSX} shows the histogram of the qubit final state, which is in approximately equal superposition (up to measurement errors) of $|0\rangle$ and  $|1\rangle$ as expected after $\sqrt{x}$ gate-operation. 

\subsection{Hadamard (H) gate}
Unlike the previous two gates, the Hadamard (H) gate is not part of the basis gates. Instead it is transpiled in terms of $\sqrt{X}$ gate and two $\pi/2$ virtual Z rotations. In our work we directly optimize the controls to implement Hadamard gate on IBM Q system. The control Hamiltonian here consists of Pauli X and Pauli Y terms. The control pulses of total length $1216$ dt $\sim$ $267$ ns  are implemented on $ibmq\_toronto$. 
Again, we use IRB to calculate the error rate for the custom H gate and the default H gate, shown in the top and middle panels of Fig.~\ref{fig:IRBH}. We find that the error rate of our custom H gate is higher compared to the default gate, which can be attributed to the longer pulse duration. The sub-optimal custom pulse is also evident from the histogram shown in the bottom panel of Fig.~\ref{fig:IRBH}, which shows that the probability of the qubit in equal-superposition-state of $|0\rangle$ and $|1\rangle$ is not exactly balanced. 
\subsection{Two-qubit gate \--- CNOT }
\begin{figure}[ht]
    \centering
    \includegraphics[width=\columnwidth]{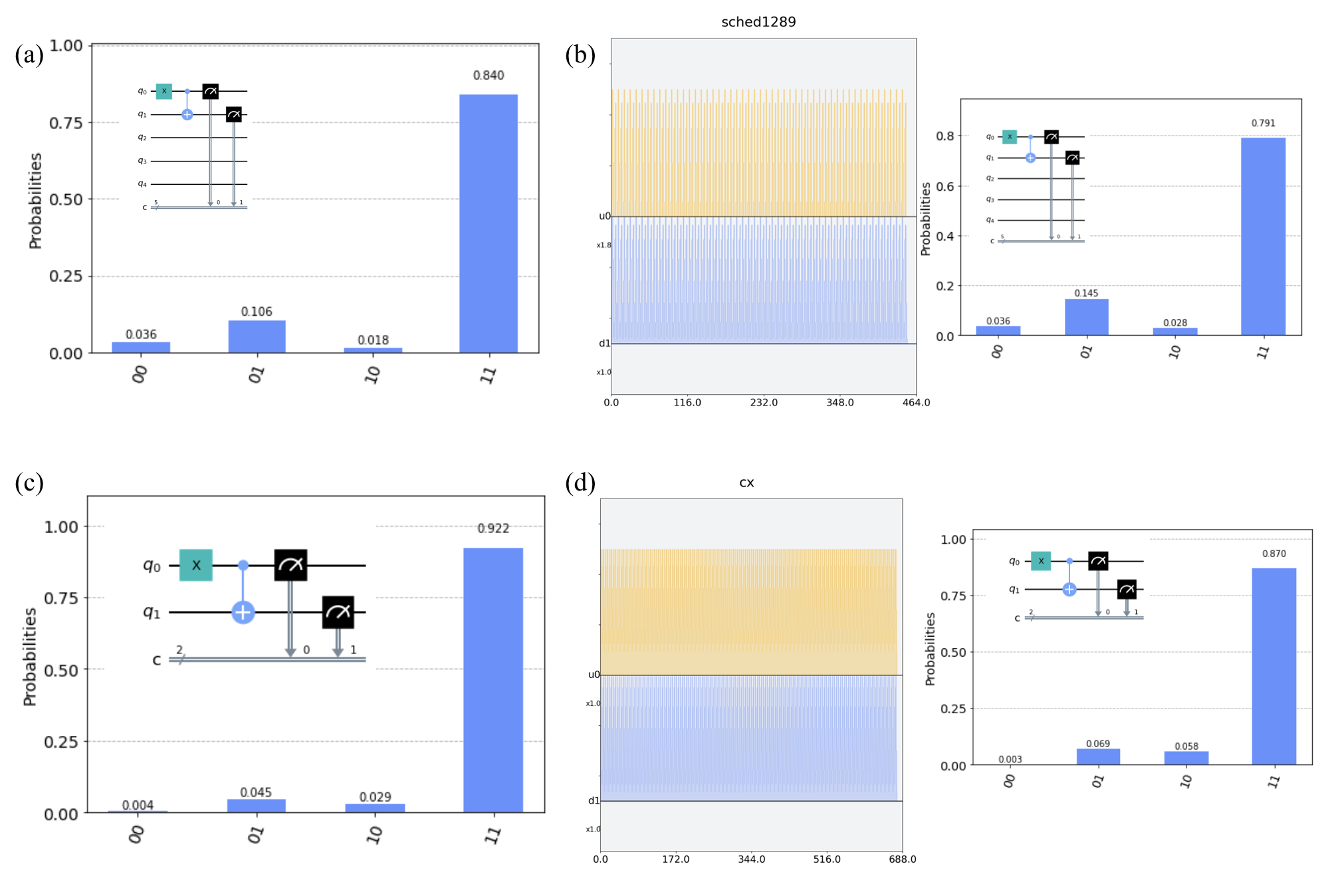}
    
     \caption{(Color online)  (a) Defaut CX gate implementation in quantum circuit and probability distribution of the quantum states on $ibmq\_boeblingen$. (b) The SINE pulse shape  and the probability distribution histogram on $ibmq\_boeblingen$. (c) Defaut CX gate implementation in quantum circuit and probability distribution of the quantum states on $ibmq\_rome$. (d) The SINE pulse shape and the probability distribution histogram on $ibmq\_rome$. }
     \label{fig:CX1}
\end{figure}

The CNOT gate is a two-qubit entangling gate which is part of the basis gate set in IBM Q. The CNOT gate is implemented by two-qubit gate known as the cross-resonance (CR) gate. The CR gate drives the target qubit through the control qubit via cross-resonance interaction.
The CR Hamiltonian as described in \cite{J.M.Chow2011} is
\begin{align}
\label{CR}
    H^{\textrm{eff}}_{\textrm{cr, drift}} &= \frac{1}{2}\Tilde{w}_{1}\sigma_{z}^{(1)}+\frac{1}{2}\Tilde{w}_{2}\sigma_{z}^{(2)}+\Omega(t)_{\textrm{R},2}\big(\sigma_{I}^{(1)}\sigma_{x}^{(2)}\big) \nonumber\\
    &+ \Omega(t)_{\textrm{R},1}\Big(\sigma_{x}^{(1)}\sigma_{I}^{(2)}+\frac{J}{\Delta_{12}}\sigma_{z}^{(1)}\sigma_{x}^{(2)}\Big).
\end{align}
From Eq. ~\ref{CR}, the control terms are $\sigma_{x}^{(1)}\sigma_{I}^{(2)},\sigma_{I}^{(1)}\sigma_{x}^{(2)},\sigma_{z}^{(1)}\sigma_{x}^{(2)}$. 
We first show our results for the ``SINE" input pulse shape implemented in the QuTiP optimizer. These results were executed on now retired $ibmq\_boeblingen$ and $ibmq\_rome$ systems. At the time of running our optimized pulses Qiskit had not released interleaved-randomized benchmarking. 
We tested our pulses by implementing it in a quantum circuit, and plot the probability distribution of the output states in Fig.~\ref{fig:CX1}. On $ibmq\_boeblingen$ with the optimized pulses the probability of getting the output state $|11\rangle$ is $79\%$, while on $ibmq\_rome$ the probability is $87\%$, both of which offer little to none improvement over the default CX gate. 

\begin{figure}[htbp]
    \centering
    \includegraphics[width=0.9\columnwidth]{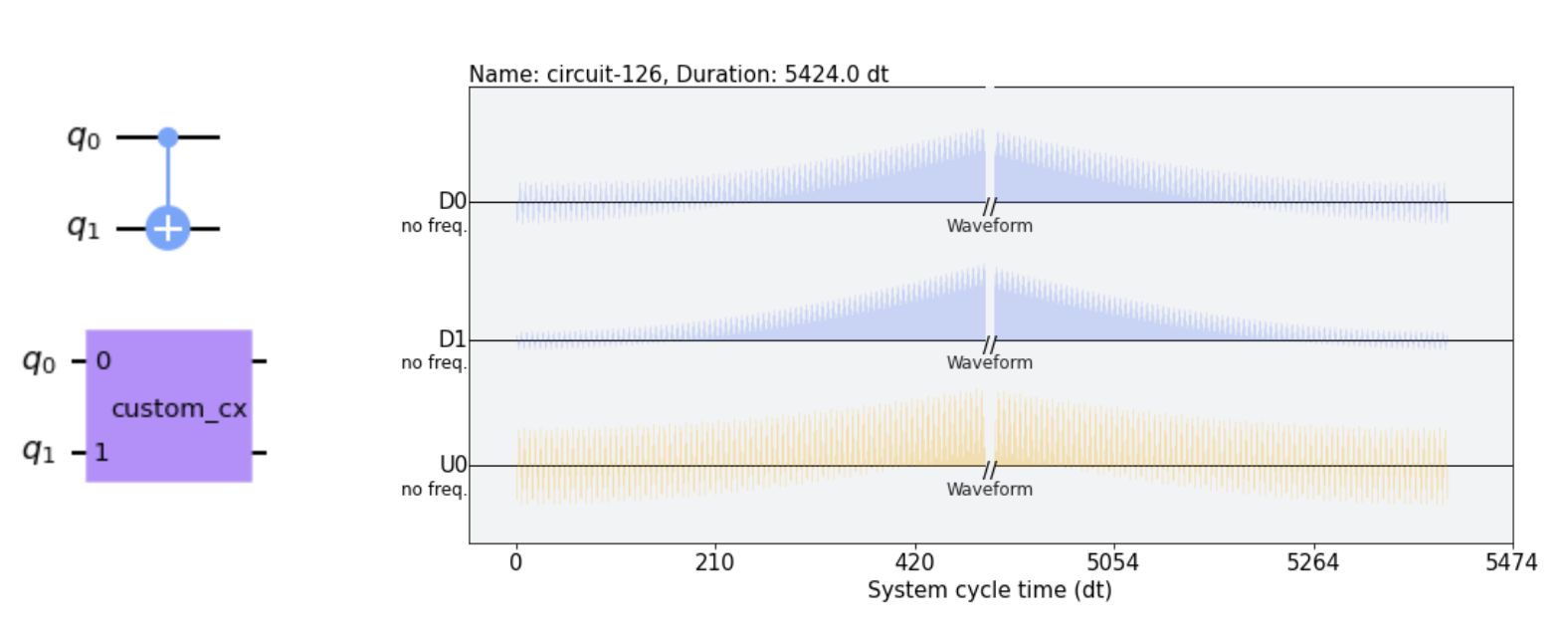}
    \caption{(Color online)The control pulse implemented on $ibmq\_montreal$ device. The pulse shape is Gaussian square shape. D0 and D1 are the DriveChannels for qubit 0 and qubit 1 respectively. U0 is the control channel for qubit 0 as the control qubit, and qubit 1 as target qubit.  The custom pulses are cast into custom gate and implemented in the quantum circuit. The default CNOT gate (CX) is replaced by our optimized CX gate, which is confirmed in the transpiling process.}
    \label{fig:CX_pulse}
\end{figure}

\begin{figure}[ht]
\centering
   \includegraphics[width=0.75\columnwidth]{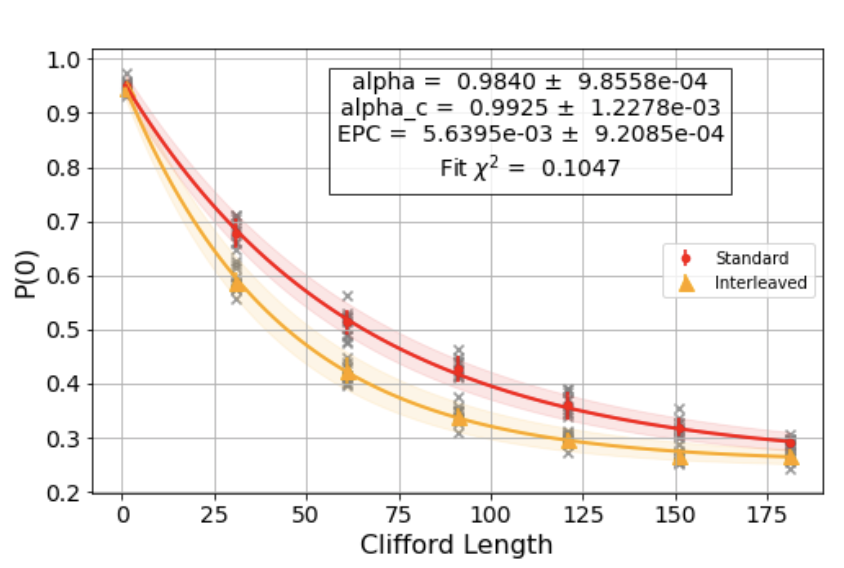}
   \includegraphics[width=0.75\columnwidth]{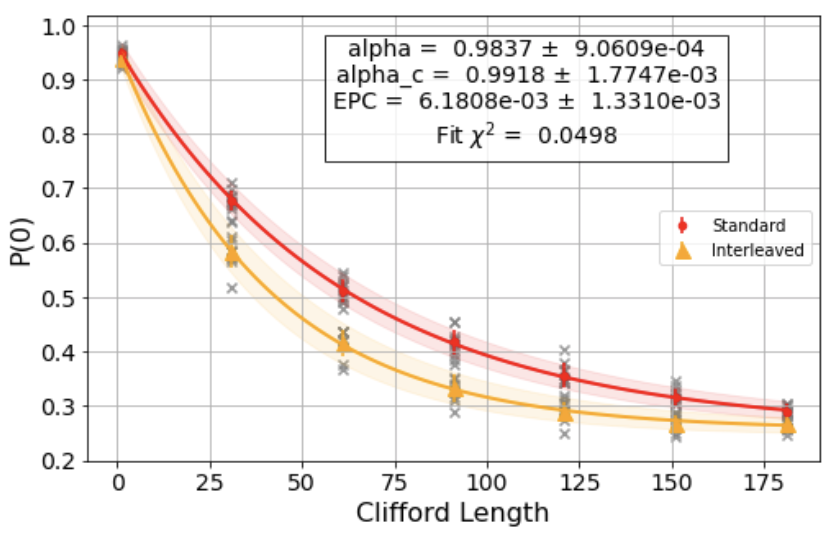}
\caption{(Color online) (Top) IRB result for the custom CX gate obtained from optimized pulse controls. The error rate here is  $(5.6\pm0.9)\times10^{-3}$. (Bottom) IRB result for the default CX gate as implemented in Qiskit. The error rate here is $(6.2\pm1.3)\times10^{-3}$. }
\label{fig:IRB_CX}
\end{figure}

Instead of first optimizing CR gate and then inserting single qubit rotations, we directly solve for CNOT gate and implement on the $ibmq\_montreal$ system for another set of control pulses with the Gaussian squared pulse shape as the input to the QuTiP solver. 
The control pulses are shown in Fig~\ref{fig:CX_pulse}.
The IRB results for the optimized CX gate are shown in Fig.~\ref{fig:IRB_CX}.
From our results we see that the average error per gate with the custom CX gate is almost the same as the default CX gate, with the custom CX gate error being only $8\%$ lower than the default CX gate. 

\section{Discussion}
The properties of an IBM Q system are dynamic and need to be updated at system calibration time. According to the information provided by the IBM Q documentation this occurs at least once over a 24-hour period. These system properties include the qubit frequency, readout error, values of $T_{1}$ and $T_{2}$. Hence, this drifting of qubit properties can lead to fluctuations in the performance of qubits over time.  
We carried out two sets of experiments to study the impact of time on our results. For the first set of experiments we optimized the pulses only once and tested the optimized pulses (for different quantum gates) on IBM Q devices on different days. 

In the second set we took into account the frequent calibration of device parameters and optimized the pulses everyday. We then executed the quantum circuit containing the corresponding day-specific pulse gate on the IBM Q device and measured the probability distribution histograms similar to those presented in Section~\ref{sec:results}. In both experiments we noted some variations in the performance of the respective gates on certain days.
We also compared these measurement results against the results of IRB and noted that the gate errors calculated by IRB were consistently low for different days, which contradicts the histogram results. This seems to indicate that IRB results are less sensitive to the device fluctuations. More detailed discussions of this issue will be presented in a separate paper. 

Additionally, from our results we see that there is only small improvement in overall gate fidelity. One of the reasons is that even though we implement single qubit gates we cannot overlook its interaction with the neighboring qubits. Therefore, for a more accurate model to input into the optimal control protocol, we need to include the qubit-qubit interaction terms. However, this does not guarantee that we would get an optimal solution, because as the complexity increases the optimal control algorithm converges slowly and is highly likely to get caught in a local minimum. 

In the case of two-qubit gates one of the major issues is the uncertainty in the Hamiltonian. There exist extra interaction terms in addition to the classical cross-talk, which could lower the fidelity of the CNOT gate. Understanding the effect of cross-talk channels on the control and target qubit dynamics is an active area of research. For multi-qubit gates it has been shown \cite{FirstPrinc2020} that higher energy levels also have an impact on the system dynamics.

We also find that the gate duration of the optimized pulses can have an effect on the average gate errors as measured by IRB. We repeated the experiments described in Section~\ref{sec:results} with shorter-duration pulses (except for the CX gate), and found that we could sometimes further improve the average gate errors. The error rates as measured by IRB for all the experiments we conducted on the IBM Q system are summarized in Table~\ref{tab:IRB}. It is worth noting that while we did not see any improvement for the custom H gate with a long 162-ns duration, at a much short duration of 31 ns, we were able to improve the average gate error with the custom gate. We have not tried a shorter gate duration for the CX gate, but will do so in the future. 

\begin{table}[htbp]
\caption{Comparison of error rate per gate with and without optimized custom pulses as measured by interleaved randomized benchmarking on IBM Q devices for different gate-durations. The default gate duration is fixed at 32 ns. Results for the X, $\sqrt{x}$, and CX gates were obtained on the $ibmq\_montreal$ system. Results for the H gate were obtained on the $ibmq\_toronto$ system.}
\begin{center}
\begin{tabular}{|c|c|c|c|c|}
\hline
\textbf{Gate}& \textbf{Duration} & \multicolumn{2}{|c|}{\textbf{IRB error rate} ($\times10^{-4}$)} & \textbf{Improvement} \\
\cline{3-4} 
\textbf{} & & \textbf{\textit{custom}}& \textbf{\textit{default}}& \textbf{} \\
\hline
X & $105$ ns & $2.0(5)$ & $2.8(5)$ & $29\%$ \\ 
X & $56$ ns  & $1.4(1.1)$  & $2.8(5)$ & $50\%$  \\
\hline
$\sqrt{x}$ & $162$ ns & $2.4(8)$ & $6.5(1.4)$ & $63\%$ \\
$\sqrt{x}$ & $31$ ns  & $4.1(2)$  & $6.5(1.4)$ & $36\%$ \\
\hline
H & $267$ ns & $26(4)$ & $5.0(7)$ & - \\
H & $28$ ns & $3.1(1.3)$ & $5.0(7)$ & $39\%$ \\
\hline
 CX & $1193$ ns & $56(9)$ & $62(13)$ & $10\%$ \\
 \hline
\end{tabular}
\label{tab:IRB}
\end{center}
\end{table}

\section{Conclusion and Outlook}

We have described our methods to improve quantum gate fidelity by optimizing the control pulse that drives the quantum system. We were able to optimize control pulses for quantum gates on superconducting quantum systems using the gradient-based optimization algorithms. We executed these pulses successfully on the IBM Q hardware and verified our results by plotting the output probability histograms and performing interleaved randomized benchamrking of our pulse-optimized gates. While IRB gives some insight into the general gate fidelity, it does not provide quantitative information on how much improvement we can expect for an actual quantum algorithm. Our next step is to apply these custom gates to more complex quantum circuits, and see how much improvement we can obtain on the accuracy of the results. It will also be interesting to see if such optimization techniques will work on other quantum architectures. 

\section*{Acknowledgments}
This research used resources of the Oak Ridge Leadership Computing Facility, which is a DOE Office of Science User Facility supported under Contract DE-AC05-00OR22725, and resources provided by the Scientific Data and Computing Center at Brookhaven National Laboratory. We thank the following colleagues for stimulating and fruitful discussions: Ning Bao, Yen-Chi Chen, Yanzhu Chen, Robert Kosut, Herschel A. Rabitz, Hubertus van Dam, Tzu-Chieh Wei,  Hongye Yu, and Yusheng Zhao. We thank Boxi Li and Alex Picthford for useful discussions on QuTiP. We acknowledge the use of IBM Quantum services for this work. The views expressed are those of the authors, and do not reflect the official policy or position of IBM or the IBM Quantum team. We would also like to thank the IBM Q support team for promptly answering our questions about Qiskit. 
 \bibliographystyle{ieeetr}
\bibliography{main}

\begin{thebibliography}{10}

\bibitem{Barnes2020}
D.~Buterakos, S.~Das~Sarma, and E.~Barnes, ``Geometrical formalism for
  dynamically corrected gates in multiqubit systems,'' {\em PRX Quantum},
  vol.~2, p.~010341, Mar 2021.

\bibitem{ESW}
P.~da~Silva, P.~S. Rouchon, P.~Silveira, and H.~Bessa, ``A fixed point
  algorithm for improving fidelity of quantum gates,'' {\em ESAIM: COCV},
  vol.~27, p.~S9, 2021.

\bibitem{CONG20209220}
S.~Cong, L.~Zhou, and F.~Meng, ``Lyapunov-based unified control method for
  closed quantum systems,'' {\em Journal of the Franklin Institute}, vol.~357,
  no.~14, pp.~9220--9247, 2020.

\bibitem{Bhole2015}
G.~Bhole, V.~S. Anjusha, and T.~S. Mahesh, ``Steering quantum dynamics via
  bang-bang control: Implementing optimal fixed-point quantum search
  algorithm,'' {\em Phys. Rev. A}, vol.~93, p.~042339, Apr 2016.

\bibitem{Krotov}
M.~H. Goerz, D.~Basilewitsch, F.~Gago-Encinas, M.~G. Krauss, K.~P. Horn, D.~M.
  Reich, and C.~P. Koch, ``{Krotov: A Python implementation of Krotov's method
  for quantum optimal control},'' {\em SciPost Phys.}, vol.~7, p.~80, 2019.

\bibitem{GRAPE1}
N.~Khaneja, T.~Reiss, C.~Kehlet, T.~Schulte-Herbrüggen, and S.~J. Glaser,
  ``Optimal control of coupled spin dynamics: design of nmr pulse sequences by
  gradient ascent algorithms,'' {\em Journal of Magnetic Resonance}, vol.~172,
  no.~2, pp.~296--305, 2005.

\bibitem{CRAB}
T.~Caneva, T.~Calarco, and S.~Montangero, ``Chopped random-basis quantum
  optimization,'' {\em Phys. Rev. A}, vol.~84, p.~022326, Aug 2011.

\bibitem{GOAT}
S.~Machnes, E.~Ass\'emat, D.~Tannor, and F.~K. Wilhelm, ``Tunable, flexible,
  and efficient optimization of control pulses for practical qubits,'' {\em
  Phys. Rev. Lett.}, vol.~120, p.~150401, Apr 2018.

\bibitem{c3-optimize}
N.~Wittler, F.~Roy, K.~Pack, M.~Werninghaus, A.~S. Roy, D.~J. Egger, S.~Filipp,
  F.~K. Wilhelm, and S.~Machnes, ``Integrated tool set for control,
  calibration, and characterization of quantum devices applied to
  superconducting qubits,'' {\em Phys. Rev. Applied}, vol.~15, p.~034080, Mar
  2021.

\bibitem{qopt2021}
H.~B. Julian D.~Teske, Pascal~Cerfontaine, ``qopt: An experiment-oriented qubit
  simulation and quantum optimal control package,'' {\em arXiv:2110.05873}, Oct
  2021.

\bibitem{He2021}
R.-H. He, R.~Wang, S.-S. Nie, J.~Wu, J.-H. Zhang, and Z.-M. Wang, ``Deep
  reinforcement learning for universal quantum state preparation via dynamic
  pulse control,'' {\em EPJ Quantum Technology}, vol.~8, no.~29,
  pp.~2196--0763, 2021.

\bibitem{qctrl2021}
Y.~Baum, M.~Amico, S.~Howell, M.~Hush, M.~Liuzzi, P.~Mundada, T.~Merkh, A.~R.
  Carvalho, and M.~J. Biercuk, ``Experimental deep reinforcement learning for
  error-robust gate-set design on a superconducting quantum computer,'' {\em
  PRX Quantum}, vol.~2, p.~040324, Nov 2021.

\bibitem{supqubit2017}
J.~J.M.Gambetta and M.Steffen, ``Building logical qubits in a superconducting
  quantum computing system,'' {\em npj Quantum Inf}, vol.~3, January 2017.

\bibitem{Krantz2019}
P.~Krantz, M.~Kjaergaard, F.~Yan, T.~P. Orlando, S.~Gustavsson, and W.~D.
  Oliver, ``A quantum engineer's guide to superconducting qubits,'' {\em
  Applied Physics Reviews}, vol.~6, no.~2, p.~021318, 2019.

\bibitem{Qiskitpulse}
T.~Alexander, N.~Kanazawa, D.~J. Egger, L.~Capelluto, C.~J. Wood,
  A.~Javadi-Abhari, and D.~C. McKay, ``Qiskit pulse: programming quantum
  computers through the cloud with pulses,'' {\em Quantum Science and
  Technology}, vol.~5, p.~044006, aug 2020.

\bibitem{JOHANSSON_QuTiP}
J.~Johansson, P.~Nation, and F.~Nori, ``Qutip 2: A python framework for the
  dynamics of open quantum systems,'' {\em Computer Physics Communications},
  vol.~184, no.~4, pp.~1234--1240, 2013.

\bibitem{Riaz2019}
B.~Riaz, C.~Shuang, and S.~Qamar, ``Optimal control methods for quantum gate
  preparation: A comparative study,'' {\em Quantum Information Processing},
  vol.~18, p.~1–26, apr 2019.

\bibitem{GRAPE2}
F.~F. Floether, P.~de~Fouquieres, and S.~G. Schirmer, ``Robust quantum gates
  for open systems via optimal control: Markovian versus non-markovian
  dynamics,'' {\em New Journal of Physics}, vol.~14, p.~073023, jul 2012.

\bibitem{SPSA}
J.~Spall, ``Implementation of the simultaneous perturbation algorithm for
  stochastic optimization,'' {\em IEEE Transactions on Aerospace and Electronic
  Systems}, vol.~34, no.~3, pp.~817--823, 1998.

\bibitem{McKay2018}
D.~C. McKay, T.~Alexander, L.~Bello, M.~J. Biercuk, L.~Bishop, J.~Chen, J.~M.
  Chow, A.~D. Córcoles, D.~Egger, S.~Filipp, J.~Gomez, M.~Hush,
  A.~Javadi-Abhari, D.~Moreda, P.~Nation, B.~Paulovicks, E.~Winston, C.~J.
  Wood, J.~Wootton, and J.~M. Gambetta, ``Qiskit backend specifications for
  openqasm and openpulse experiments,'' 2018.

\bibitem{RB}
E.~Knill, D.~Leibfried, R.~Reichle, J.~Britton, R.~B. Blakestad, J.~D. Jost,
  C.~Langer, R.~Ozeri, S.~Seidelin, and D.~J. Wineland, ``Randomized
  benchmarking of quantum gates,'' {\em Phys. Rev. A}, vol.~77, p.~012307, Jan
  2008.

\bibitem{IRB}
E.~Magesan, J.~M. Gambetta, B.~R. Johnson, C.~A. Ryan, J.~M. Chow, S.~T.
  Merkel, M.~P. da~Silva, G.~A. Keefe, M.~B. Rothwell, T.~A. Ohki, M.~B.
  Ketchen, and M.~Steffen, ``Efficient measurement of quantum gate error by
  interleaved randomized benchmarking,'' {\em Phys. Rev. Lett.}, vol.~109,
  p.~080505, Aug 2012.

\bibitem{IBMQ}
I.~Quantum., 2021.

\bibitem{Qiskit-Textbook}
A.~Asfaw, A.~Corcoles, L.~Bello, Y.~Ben-Haim, M.~Bozzo-Rey, S.~Bravyi,
  N.~Bronn, L.~Capelluto, A.~C. Vazquez, J.~Ceroni, R.~Chen, A.~Frisch,
  J.~Gambetta, S.~Garion, L.~Gil, S.~D. L.~P. Gonzalez, F.~Harkins,
  T.~Imamichi, H.~Kang, A.~h.~Karamlou, R.~Loredo, D.~McKay, A.~Mezzacapo,
  Z.~Minev, R.~Movassagh, G.~Nannicini, P.~Nation, A.~Phan, M.~Pistoia,
  A.~Rattew, J.~Schaefer, J.~Shabani, J.~Smolin, J.~Stenger, K.~Temme, M.~Tod,
  S.~Wood, and J.~Wootton., ``Learn quantum computation using qiskit,'' 2020.

\bibitem{J.M.Chow2011}
J.~M. Chow, A.~D. C\'orcoles, J.~M. Gambetta, C.~Rigetti, B.~R. Johnson, J.~A.
  Smolin, J.~R. Rozen, G.~A. Keefe, M.~B. Rothwell, M.~B. Ketchen, and
  M.~Steffen, ``Simple all-microwave entangling gate for fixed-frequency
  superconducting qubits,'' {\em Phys. Rev. Lett.}, vol.~107, p.~080502, Aug
  2011.

\bibitem{FirstPrinc2020}
M.~Malekakhlagh, E.~Magesan, and D.~C. McKay, ``First-principles analysis of
  cross-resonance gate operation,'' {\em Phys. Rev. A}, vol.~102, p.~042605,
  Oct 2020.

\end{thebibliography}
\end{document}